\documentclass[conference]{IEEEtran}
\IEEEoverridecommandlockouts
\usepackage{cite}
\usepackage{amsmath,amssymb,amsfonts}
\usepackage{algorithmic}
\usepackage{graphicx}
\usepackage{textcomp}
\usepackage{xcolor}

\usepackage[]{bm}
\usepackage{multirow}
\usepackage{booktabs}
\usepackage{rotating}
\usepackage{subcaption}
\usepackage[bookmarks=false]{hyperref}

\def\BibTeX{{\rm B\kern-.05em{\sc i\kern-.025em b}\kern-.08em
    T\kern-.1667em\lower.7ex\hbox{E}\kern-.125emX}}
\begin{document}

\title{Approximate Manifold Defense Against Multiple Adversarial Perturbations\\
}

\author{\IEEEauthorblockN{Jay Nandy}
\IEEEauthorblockA{\textit{School of Computing} \\
\textit{National University of Singapore} \\
jaynandy@comp.nus.edu.sg}
\and
\IEEEauthorblockN{Wynne Hsu}
\IEEEauthorblockA{\textit{School of Computing} \\
\textit{National University of Singapore} \\
whsu@comp.nus.edu.sg}
\and
\IEEEauthorblockN{Mong Li Lee}
\IEEEauthorblockA{\textit{School of Computing} \\
\textit{National University of Singapore} \\
leeml@comp.nus.edu.sg}
}

\maketitle

\begin{abstract}
Existing defenses against adversarial attacks are typically tailored to a specific perturbation type. 
Using adversarial training to defend against multiple types of perturbation requires expensive adversarial examples from different perturbation types at each training step.
In contrast, manifold-based defense incorporates a generative network to project an input sample onto the clean data manifold. 
This approach eliminates the need to generate expensive adversarial examples while achieving robustness against multiple perturbation types.
However, the success of this approach relies on whether the generative network can capture the complete clean data manifold, which remains an open problem for complex input domain.
In this work, we devise an approximate manifold defense mechanism, called RBF-CNN, for image classification. 
Instead of capturing the complete data manifold, we use an RBF layer to learn the density of small image patches.
RBF-CNN also utilizes a reconstruction layer that mitigates any minor adversarial perturbations.
Further, incorporating our proposed reconstruction process for training improves the adversarial robustness of our RBF-CNN models.
Experiment results on MNIST and CIFAR-10 datasets indicate that RBF-CNN offers robustness for multiple perturbations without the need for expensive adversarial training.

\end{abstract}

\begin{IEEEkeywords}
Deep Learning, Adversarial attack, Robustness, Image classification, RBF filter, EM algorithm.
\end{IEEEkeywords}

\section{Introduction}
\label{chap4:intro}
Despite the impeccable success of deep neural network (DNN)-based models in various applications, there is a growing awareness of their vulnerability against adversarial attacks \cite{advStart_iclr_2014,fgsm_iclr_2015}. 
An adversary deliberately introduces minor perturbations that mislead the networks to produce wrong predictions for the perceptually identical inputs.
The adversarial vulnerability of DNN models has led to concern about the safety and reliability of these models for real-world applications \cite{traffic_euroSP_2016,spam_icassp_2013,ad_arxiv_2018}. 

Several methods have been proposed to improve the robustness of DNN models against adversarial attacks.
Two of the most successful defense frameworks against adversarial attacks are adversarial training and randomized smoothing.
The \textit{adversarial training} mechanism trains a model using adversarial examples of a specific $\ell_p$ perturbation type to achieve robustness for that perturbation type \cite{madry_iclr_2018,treadAdv_icml_2019}.
This training process is expensive as it requires the generation of adversarial examples at each training iteration.
Tram{\`e}r and Boneh (2019) \cite{multiple_advTrain_2019} demonstrates that adversarial training can achieve robustness for multiple perturbation types only by incorporating different types of adversaries for training. 
However, this raises the question of how many perturbation types one should include for training?

\textit{Randomized smoothing} technique introduces run-time randomization that evaluates multiple noisy copies of a test image and returns the most probable class as their final prediction \cite{certifyNoise_icml_2019,certifyNoise_nips_2019}. 
This framework provides certified robustness for minor-$\ell_2$ perturbation.
However, it offers no guarantee for other perturbation types.

In contrast, \textit{manifold-based defenses} incorporate a generative network to project an input image into the clean data manifold and have the potential to achieve robustness for multiple perturbation types
\cite{defenseGAN_iclr_2018,vaeDef_aaai_2019,mnist_multiple_iclr2019}.
However, due to the limited capacities of existing generative models, they often fail to capture the complete data manifold for complex image domains and project the adversarial images into the clean data manifold.
To date, Schott et al. (2019) \cite{mnist_multiple_iclr2019} provide the only effective manifold-based defense to achieve robustness for MNIST.
Even then, they cannot train a single robust classifier for all $\ell_{p\geq 1}$ perturbation types.

In this work, we propose a novel manifold-based defense framework for image classification that can scale to complex data manifold and achieve robustness against any minor for $\ell_{p\geq 1}$ perturbation.
 Given an image, $\bm{x}$ and some perturbation bound $\epsilon_p$, for all $p\geq 1$,
the goal is to build a classifier $\mathcal{F}$ such that the  prediction remains unchanged, that is, 
\begin{equation} \small
\mathcal{F}(\bm{x}) = \mathcal{F}(\bm{x +  \delta}), 
\qquad ||\bm{\delta}||_{p} < \epsilon_p, \qquad \forall ~ p \geq 1
\end{equation}

To this end,  we devise an approximate manifold defense mechanism called  \textit{RBF-CNN} that can achieve robustness for any minor $\ell_{p\geq 1}$ perturbation.
Our RBF-CNN models consist of a radial basis function (RBF) layer and a reconstruction layer followed by a convolutional neural network (CNN) image classifier.
An RBF layer consists of RBF filters and is utilized as a generative structure to capture the density of small image patches, instead of the distribution of full-sized training images \cite{nsn_icip_2018}.
Each RBF filter acts as a template matching function by producing similar match scores for two similar patterns in any $\ell_{p\geq 1}$-norms \cite{fgsm_iclr_2015}.
We show that our reconstruction layer utilizes this property to mitigate any minor perturbation in any $\ell_{p\geq 1}$-norm.
Further, incorporating the reconstruction process for training allows us to improve the adversarial robustness of our RBF-CNN models.

Experimental results on MNIST and CIFAR-10  demonstrate that our RBF-CNN models achieve robustness against all $\ell_1$, $\ell_2$ and $\ell_{\infty}$ bounded adversarial attacks and provide certified robustness for $\ell_2$ bounded perturbations. 
The saliency maps produced by RBF-CNN models conform to human interpretation and thereby ensure that our RBF and reconstruction layers do not cause gradient obfuscation \cite{robustnessDrops_iclr_2018}.
Therefore, our framework does not provide a false sense of robustness by resisting attackers to generate optimal adversarial examples \cite{obfuscated_icml_2018}. 
Further, these interpretable saliency maps provide visual evidence of robustness against adversarial attacks \cite{saliency_icml_2019}.
To the best of our knowledge, RBF-CNN is the first manifold based defense framework to achieve robustness for any $\ell_{p\geq 1}$ perturbations and provides a desirable trade-off between robustness vs. accuracy at run-time.

\section{Related work}
\label{chap4:related}
Several defense models have been proposed to improve the robustness of DNN classifiers against adversarial perturbations.
Among the existing defense frameworks, adversarial training provides the best empirical robustness against adversarial attacks. 
Madry et al. (2018) \cite{madry_iclr_2018} achieve $\ell_{\infty}$-robustness by training the classifier using expensive $\ell_{\infty}$-bounded PGD adversaries.
This framework can be generalized to any perturbation type, and the interpretable loss-gradients indicated that the classifiers do not cause gradient masking \cite{robustnessDrops_iclr_2018}\cite{obfuscated_icml_2018}.
Recent works such as TRADES \cite{treadAdv_icml_2019}, LLR \cite{llrAdv_nips_2019} propose additional loss regularizer for adversarial training models.
Shafahi et al. (2019) \cite{freeAdv_nips_2019} propose a `free'-adversarial training to improve the training efficiency by unfolding the PGD-adversary generation step.
However, Tram{\`e}r and Boneh (2019) \cite{multiple_advTrain_2019} demonstrates that it would require the generation of adversarial examples for different perturbation types to achieve robustness for multiple-perturbation, leading to a linear increase in the training time.
Their findings have also raised concerns that adversarial training may not be the appropriate direction to improve the adversarial robustness for multiple perturbations \cite{multiple_advTrain_2019}\cite{mnist_multiple_iclr2019}.

Input transformation techniques such as feature squeezing \cite{featureSq_2017},  pixel-deflection \cite{pixelDeflection_cvpr_2018} have been proposed to heuristically alter the pixels in images. 
However, these techniques did not induce the correct amount of alteration and thus, could not improve the robustness \cite{obfuscated_icml_2018}.
Xie et al. (2018) \cite{denoiserDefense_cvpr_2018} achieve $\ell_{\infty}$-robustness  by incorporating adversarial training with their denoising network.
In contrast, randomized smoothing defenses induce the correct amount of noise to achieve certified robustness \cite{certifyDP_2018}\cite{certifyNoise_icml_2019}\cite{certifyNoise_nips_2019}. 
However, their success is limited to minor-$\ell_2$-perturbations.

Image quilting \cite{transform_iclr_2018} is another type of input transformation technique. 
It replaces the original input patches using the clean patches, selected from a large set of $50,000$ clean patches.
This defense has been broken by the adaptive BPDA attack \cite{obfuscated_icml_2018}.
Unlike the quilting technique, we reconstruct patches using a combination of patch samples, randomly drawn from a much smaller set of our RBF filters, that also inject noises to the images.
Instead of using pre-trained weights for the CNN classifiers like \cite{transform_iclr_2018}, we train our model by incorporating the reconstruction layer and achieve a robust classifier for any minor $\ell_{p\geq 1}$ perturbations.

Our proposed defense can be categorized under manifold defense strategy where generative models are applied to project the input or the hidden activations onto the (learned) data manifold.
Examples include DefenseGAN \cite{defenseGAN_iclr_2018} using generative adversarial networks \cite{gan_nips_2014}, Ape-GAN \cite{apeGAN_defense_2019} using auto-embedding GAN \cite{aeGAN_ieee_2019}, defense using Variational Autoencoder (VAE) \cite{vaeDef_aaai_2019}, PixelDefend \cite{pixeldefend_iclr_2018}.
\cite{mnist_multiple_iclr2019} train multiple VAEs for each class and chose the class with maximum-likelihood score as their prediction.
Similar ideas using non-generative models include auto-encoders to project the input images into the `known' data manifold \cite{magnet_2017,denoiserDefense_cvpr_2018}.
However, all  these defenses, except \cite{mnist_multiple_iclr2019}, are  found to be ineffective \cite{obfuscated_icml_2018}\cite{overpoweredAttack_2017}.

The key reason why these defenses fail is that they rely on their generative model(s) to completely remove adversarial perturbations.
This would require the generator to completely capture the underlying data manifold.
However, due to the limited capacities of the existing generative (or non-generative) models, they fail to efficiently capture the complete data manifold and often over-estimate or under-estimate different modes of the underlying distribution.
Further, the classifier used to classify these projected images, often remain adversarially vulnerable as before.
Hence, these models are either broken or fail to scale to complex image domains.
Here, we address these limitations by using an RBF layer
 to capture the density of smaller input image patches.
Our reconstruction process allows us to train  classifiers that
is robust against any minor $\ell_{p\geq 1}$-perturbations and remain effective on a more complex image dataset such as CIFAR-10.

The most recent work Croce and Hein (2020) \cite{lp_iclr_2020} design a regularization scheme to provide robustness for small networks for all $\ell_{p\geq 1}$ perturbations. 
However, their success remains limited for much smaller perturbation boundaries and at the cost of much reduced clean data classification accuracy.

\smallskip
\section{Proposed framework}

\begin{figure*}[t]
    \centering
    \includegraphics[width=0.9\linewidth,height=90pt]{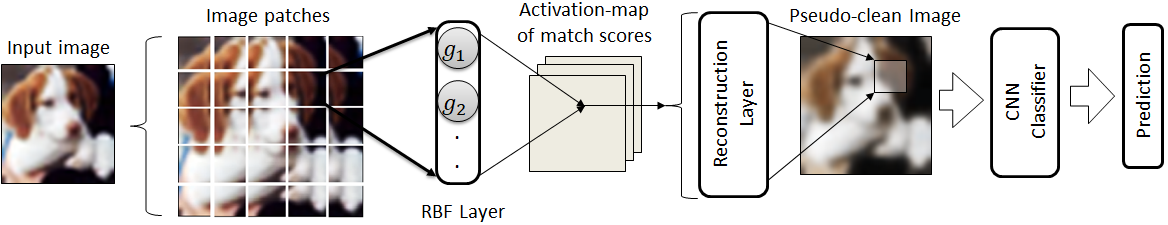}    
    \caption{ Overview of the proposed RBF-CNN framework.}
    \label{fig_overview}
\end{figure*}

Figure \ref{fig_overview} gives an overview of our proposed RBF-CNN framework. 
An input image is first passed through the RBF layer to obtain the activation maps of match scores.
The reconstruction layer then combines the patch samples, drawn from the RBF filters and use the match scores to produce pseudo-clean images for classification.

\smallskip
\noindent
\paragraph{\textbf{RBF Layer}}
An RBF layer is similar to a convolutional layer, except that the convolutional filters are replaced by RBF filters.
Given an input image ${\bm x}$, we slide a fixed-sized window to obtain the image patches $\bm{z}_i, i = \{1, 2, \cdots \}$.
For each image patch $\bm{z}$, we compute the \textit{match score} $s(\bm{z},g)$ from an RBF filter $g(\bm{\mu}, \sigma)$ as follows:
%
\begin{equation} \small
\label{GCN-sim}
s(\bm{z},g) = \log \Bigg( \frac{1}{\sqrt[]{2\pi}\sigma} 
\exp -\frac{||\bm{z} - \bm{\mu}||^2_2}{2\sigma^2} \Bigg) 
\end{equation}
where the filter-mean $\bm{\mu}$ is of the same shape as the image patch $\bm{z}$, and  $\sigma \in \mathbb{R}$ denotes the spread of the filter.

\begin{figure}[h]
	\centering
	\includegraphics[width=0.8\linewidth,height=100pt]{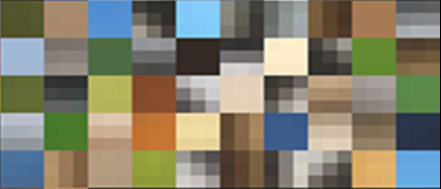}    
	\caption{Visualization of a few RBF filter means of patch size $3 \times 3$ for the CIFAR-10 classification model.}
	\label{fig_patch}\vspace*{-0.1in}
\end{figure}

As we can see in Eqn. \ref{GCN-sim}, an RBF filter produces a high match score $s(\bm{z},g)$ for all the nearby patches $\bm{z}$ such that the Euclidean distance $||\bm{z} - \bm{\mu}||^2_2$ is small.
Thus, an RBF filter $g(\bm{\mu}, \sigma)$ can be viewed as a  template matching function
where the mean $\bm{\mu}$ acts as a template to be matched with the input image patch $\bm{z}$ to produce the match score.
Figure \ref{fig_patch} visualizes a few RBF filter means of size $(3 \times 3)$ learned from clean CIFAR-10 training images.

However, learning the RBF filters, along with the CNN classifier, using the back-propagation algorithm is difficult as they may not be able to efficiently capture the density of image patches while minimizing classification loss.
We overcome this by separately training the RBF layer in an unsupervised fashion to capture the density of the image patches.
Here, we use a non-parametric variant of expectation maximization (EM) algorithm to learn the filter parameters and capture the density of the image patches \cite{nsn_icip_2018}.

The non-parametric EM is a hard clustering algorithm that automatically determines the required number of filters. 
We start with one filter  $g_1(\bm{\mu}_1,\sigma_1)$ where the filter mean $\bm{\mu}_1$ and scaling parameter $\sigma_1$ are randomly initialized. 
New filters are then created subsequently in the following iterations consisting of E and M steps.

The E-Step computes the  match score $s(\bm{z}_i,g_j)$ of the filter $g_j(\bm{\mu}_j,\sigma_j)$ for each clean training image patch $\bm{z}_i$.
We assign patches with maximum match scores to the RBF filters.
If the score is lower than some pre-determined threshold, we create a new filter for $\bm{z}_i$. 
The  M-step updates the parameters of the RBF filters using the assigned patches as follows:
\begin{equation} \small
\label{crp-M}
\bm{\mu}_j^{next} = \frac{\sum_{\bm{z}_i \in J} \bm{z}_i}{n_J}, \qquad
\sigma_j^{next} =  \sqrt {\frac{\sum_{\bm{z}_i \in J} ||\bm{z}_i - \bm{\mu}_j^{next}||^2_2}{n_J}}
\end{equation}
where $J$ is the set of patches assigned to the filter $g_j$ and $n_J$ denotes the cardinality of set $J$.

\smallskip
\smallskip
\paragraph{\textbf{Reconstruction Layer}}
Our reconstruction layer does not contain any learn-able parameters. 
It reuses the RBF filters to separately reconstruct each input patch to produce a pseudo-clean image in the following two steps.

\smallskip
\noindent
\textit{Step 1. Compute weight-vector from the match scores. }\\
For each input patch $\bm{z}$, we obtain the match scores $ s(\bm{z},g_j)~\forall j= 1, 2, \cdots $ and apply sigmoid activation to obtain the vector $\bm{v}$.
The dynamic range of $\bm{v}$ is  increased by applying an element-wise exponential function $\exp ({\beta_1 \bm{v}})$ where $\beta_1$ is a hyper-parameter.
This is followed by a normalization operation to obtain the weight vectors $\bm{w}$.

\noindent
\textit{Step 2. Draw samples from RBF filters. }\\
We draw samples from all the RBF filters $g_j(\bm\mu_j, \sigma_j)$ as:
\begin{equation} \small
\label{eq:sample}
\bm{\tilde{z}}_j \sim \mathcal{N}(\bm\mu_j, \sigma_j \beta_2I) 
= \bm\mu_j + \mathcal{N}(0, \sigma_j \beta_2I)
\end{equation}
where $\beta_2$ is a hyper-parameter that controls the amount of noise to be added to the  patch  $\bm{\tilde{z}}_j$, and $I$ is the identity matrix.

We reconstruct the patch as the weighted sum of these samples $\bm{w}^T\bm{Z}$.
The reconstructed image is obtained by reconstructing all the original input patches of image $\bm{x}$ and stitching them together by averaging the overlapping regions.
Note that 
RBF-CNN is a randomized framework where the noise is injected during the sampling process in the reconstruction layer.
This  differs from the existing randomization smoothing techniques, where noise is 
injected in the input layer \cite{certifyNoise_icml_2019}\cite{certifyDP_2018}.

The noise-level hyper-parameter $\beta_2$ controls the trade-off between robustness and accuracy during inference.
We can pick an appropriate value for $\beta_2$ to obtain the desired level of robustness at run-time. 
To achieve high accuracy, we can set $\beta_2 = 0$ and obtain $\bm{z}_j = \bm\mu_j$ in Eqn. \ref{eq:sample}.
In our experiments, we evaluate the robustness of RBF-CNN for different values of $\beta_2$, and demonstrate that our proposed framework remains robust for all $\ell_1$, $\ell_2$ and $\ell_{\infty}$ perturbations even for $\beta_2 = 0$.
Alternatively, as we choose a higher value for $\beta_2$, the sample space of image patches for reconstruction is increased, thus improving the robustness of the classifier.

\begin{figure*}[th]
	\center
	\includegraphics[width=0.9\linewidth,height=130pt]{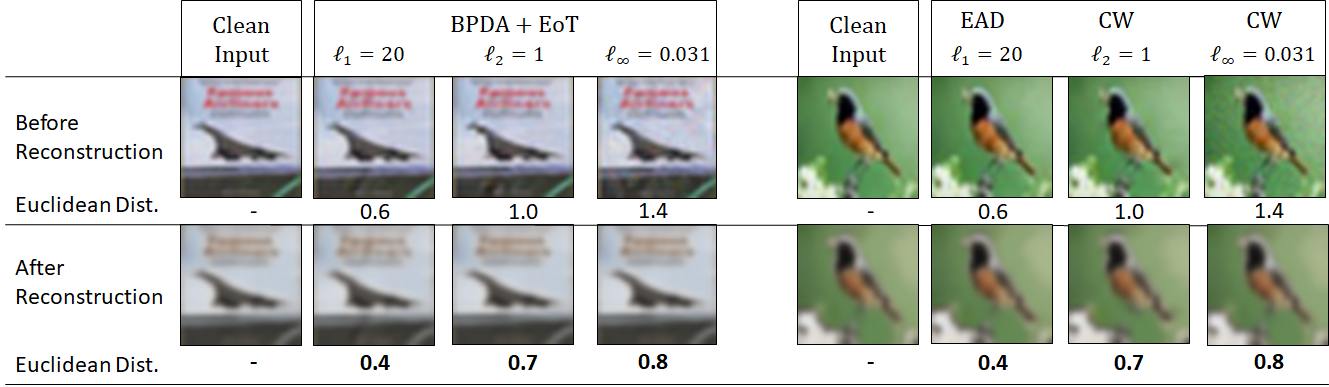}
	\caption{Visualizing the effect of our reconstruction process to mitigate minor perturbations in any $\ell_{p\geq}1$ norm.}
	\label{fig:theorem}
\end{figure*}

\smallskip
\paragraph{\textbf{Classification}}
The RBF and reconstruction layers effectively mitigate any minor $\ell_{p\geq 1}$ perturbations before feeding the images into the CNN classifier.
Since the reconstruction process incorporates randomization, we obtain the final prediction as an average of $m$ different runs as:
\begin{equation} \small
argmax \sum_{i=1}^{m} rCNN_{\beta_2}({\bm x})
\end{equation}

In our experiments, we use $m=10$. 
In other words, for each test image, we create a batch of size $m$ for the same image and execute them in parallel and obtain the final prediction by averaging these $m$ prediction.

\subsection{\textbf{Analysis of the Reconstruction Process}}
An RBF filter acts as a template matching function that computes the match score as a function of the Euclidean distance between an input patch and the filter mean 
(see Eqn. \ref{GCN-sim}).
Hence, as an attacker chooses minor $\ell_{p\geq 1}$ adversarial perturbations, the match scores produced by the RBF filters change minimally.
The reconstruction layer then combines the samples drawn from the RBF filters using the normalized match scores as the weight vectors to reconstruct pseudo-clean images, thereby mitigating minor perturbations.

~\\
\textbf{Claim 1.}
\textit{Reconstruction process mitigates the effect of minor $\ell_{p\geq 1}$ bounded adversarial perturbations and produces pseudo-clean images for classification.}

~\\
\textbf{Proof.}
We provide the proof for only $\ell_{\infty}$-norm.
The proof for other $\ell_{p\geq 1}$ can be obtained similarly.
Let $\bm{x'}$ be an adversarial image, obtained from a clean image $\bm{x}$ by modifying the pixels as: $x'_i = x_i + \delta_i$.
Let ${\bm z'}$ be the patch obtained from the adversarial image $\bm{x'}$.
Suppose $g({\bm \mu}, \sigma)$ is an RBF filter that produces a match score, $s({\bm z},g)$ for a clean patch ${\bm z}$ of ${\bm x}$ (recall Eqn. \ref{GCN-sim}).
Then we can express the match score $s({\bm z}',g)$ as:
\begin{equation} \small
\begin{split}
\label{eq:diff}
s({\bm z}',g) 
&= s({\bm z},g) - \frac{\sum_i \delta_i^2}{2\sigma^2} 
                - \frac{\sum_i \delta_i(x_i-\mu_i)}{\sigma^2}
\end{split}
\end{equation}
We first establish that   $s({\bm z}',g) - s({\bm z},g)$ is bounded.
For $\ell_{\infty}$ perturbations, we can bound $\sum_i \delta_i(x_i-\mu_i)$ as follows:
\begin{equation} \small
\begin{split}
\label{eq:inf_1}
\Big|\sum_i \delta_i(x_i- &\mu_i)\Big| \leq \delta_{max} \sum_i |(x_i-\mu_i)| \\
& = \delta_{max} ||\bm{z}-\bm{\mu}||_1
\leq \delta_{max} \sqrt{n_z} ||\bm{z}-\bm{\mu}||_2,
\end{split}
\end{equation}
where $\delta_{max} = ||{\bm \delta}||_{\infty}$ and $n_{\bm z}$ is the number of pixels in $\bm{z}$.

Here, we use \textit{Cauchy-Schwarz inequality} to get
$||\bm{z}-\bm{\mu}||_1 \leq \sqrt{n_{\bm z}} ||\bm{z}-\bm{\mu}||_2$.
Since 
 $\frac{\sum_i \delta_i^2}{2\sigma^2}>0$, we combine Eqn. \ref{eq:diff} and Eqn. \ref{eq:inf_1} to obtain the bound as:
\begin{equation} \small
\begin{split}
\label{eq:inf_bound} 
 - \frac{n_{\bm z} \delta_{max}^2}{2\sigma^2} 
                &- \frac{\delta_{max} \sqrt{n_{\bm z}} ||\bm{z}-\bm{\mu}||_2}{\sigma^2} \\
&\leq s({\bm z}',g)- s({\bm z},g)
<  \frac{\delta_{max} \sqrt{n_{\bm z}} ||\bm{z}-\bm{\mu}||_2}{\sigma^2}
\end{split}
\end{equation}

We show that this is insignificant to mitigate any minor $\ell_{\infty}$-perturbation.
Since $\delta_{max}$ is small, the term $\frac{\sum_i \delta_{max}^2}{2\sigma^2} \rightarrow 0$ in  Eqn. \ref{eq:inf_bound}.
If $g({\bm \mu}, \sigma)$ produces a high match score for patch ${\bm z}$, then $||\bm{z}-\bm{\mu}||_2$ must be small.
Hence, $\frac{\delta_{max} \sqrt{n_z} ||\bm{z}-\bm{\mu}||_2}{\sigma^2} \rightarrow 0$.
Thus, the lower bound of $s({\bm z}',g)$ of $g({\bm \mu}, \sigma)$ remains almost the same as $s({\bm z},g)$ and $g$ still produces a high score.

In contrast, a low match score for ${\bm z}$ by an RBF filter $g'(\bm{\mu}', \sigma')$ implies $||\bm{z}-\bm{\mu'}||_2$ is large (recall Eqn. \ref{GCN-sim}).
We analyze the maximum value attained by $g'(\bm{\mu}', \sigma')$ and obtain:
\begin{equation} \small
\label{eq:th_inf_bound2}
s({\bm z}',g') <
\log \frac{1}{\sqrt{2\pi}\sigma'}
- \frac{(||\bm{z-\mu'}||_2 - 2\delta_{max} \sqrt{n_{\bm z}})||\bm{z-\mu'}||_2 }{2(\sigma')^2}
\end{equation}

Hence, to achieve a high value of $s({\bm z}',g')$, the term $(||\bm{z-\mu'}||_2 - 2\delta_{max} \sqrt{n_{\bm z}})$ should be small.
However, since $\delta_{\max}$ is small and $||\bm{z}-\bm{\mu'}||_2$ is large, we have $||\bm{z-\mu'}||_2 >> 2\delta_{max} \sqrt{n_{\bm z}}$.
Hence, $g'$ still produces low match score for ${\bm z'}$.

\smallskip
Since the difference between the match scores of clean images and the corresponding adversarial images remain insignificant, the weight vectors for the reconstruction process will hardly change.
Hence, the reconstruction process mitigates their distance when reproducing images from the same set of samples, drawn from the RBF filters.
$\hfill\Box$

\smallskip
\smallskip
Figure \ref{fig:theorem} illustrates that our reconstruction process mitigates the distances between the images irrespective of all minor perturbation in any $\ell_{p\geq1}$ norm.
Hence, our reconstruction process would enforce the adversaries to choose larger perturbation bounds to circumvent the classifiers.
However, we still need to robustly classify these reconstructed images with mitigated perturbations.

\smallskip
\smallskip
\textbf{Incorporating the reconstruction process for training improves the adversarial robustness.}
 Schmidt et al. (2018) \cite{moreData_nips_2018} and  Hendrycks et al. (2019) \cite{pretrainRobust_icml_2019} have shown that adversarial robustness of a DNN classifier improves by incorporating more training images. 
On the other hand, by following our Claim 1, we argue that our reconstruction process mitigates the distance between any two images within a small neighborhood in any $\ell_{p\geq1}$ norm.
In other words, it projects the original manifold of input images into a more \textit{compact manifold} of reconstructed images.
Hence, incorporating our proposed reconstruction process for training reduces the requirement of additional images to improve the robustness of our RBF-CNN models.
Moreover, our Claim 1 implies that our reconstruction process would produce almost the same image for a set of input images in a small neighborhood of any $\ell_{p\geq1}$-norm.
Hence, by augmenting minor random noise to our training images, we can efficiently train our models to be more aware of the surroundings of the data manifold.
Our experiments on MNIST and CIFAR-10 supports this observation.

\section{Performance Study}
We carried out three sets of experiments to evaluate the robustness of our RBF-CNN models  on MNIST \cite{db_mnist} and CIFAR-10 \cite{db_cifar}\footnote{Code is available at
\href{https://github.com/jayjaynandy/RBF-CNN}{\texttt{\small //github.com/jayjaynandy/RBF-CNN.}}}.
First, we empirically evaluate the robustness against a wide range of $\ell_1$, $\ell_2$ and $\ell_{\infty}$ bounded attacks.
We demonstrate that RBF-CNN models improve certified robustness for $\ell_2$ perturbations.
Next, we visualize that our RBF-CNN models produce interpretable saliency maps to ensure that our framework does not cause gradient obfuscation \cite{robustnessDrops_iclr_2018}.
Finally, we demonstrate that our RBF-CNN models allow robustness vs. accuracy flexibility at run-time.

\smallskip
\noindent \textbf{Experimental Setup. }
We use a  $4$-layer CNN for MNIST, and VGG-16 \cite{vgg_iclr_2015} for CIFAR-10.
We train two sets of RBF-CNN models, denoted as $rCNN$ and $rCNN_+$. 
We use $3\times 3$ filters for the RBF layer and train with the non-parametric EM that automatically learns $24$ and $232$ filters for MNIST and CIFAR-10 respectively.
The CNN components of these models are trained using \textit{label smoothing}.
We use only clean images to train our $rCNN$ models. 
For $rCNN_+$ models, we use the clean images as well as noisy images, perturbed within a $\ell_{\infty}$ boundary of $0.3$ and $0.03$ for MNIST and CIFAR-10 respectively.
Here, the noises for $rCNN_+$ is sampled from an isotropic Gaussian ($\mathcal{N}(0, 0.35)$ and $\mathcal{N}(0, 0.05)$ respectively) and clipped within those predefined $\ell_{\infty}$ boundaries.
We also generate one set of PGD-adversarial examples of the training images at $25$ and $200$ epochs for MNIST and CIFAR-10 respectively.
Then we obtain the \textit{adversarial noises} by subtracting the original training images from the adversarial examples.
We randomly add this adversarial noise to the clean training images for the rest of the training epochs.

During testing,  $\beta_1$ is set to 25 for both datasets.
The hyper-parameter $\beta_2$ controls the trade-off between accuracy and robustness for the classifiers.
We do not inject any noise through the reconstruction layer during training.
For testing, we set $\beta_2$ to $1.75$ for both MNIST and CIFAR-10.
For CIFAR-10, we choose $\beta_2$ such that the classification accuracy for clean images does not drop less than $85\%$.

\subsection{Performance against Adversarial Attacks}
The \textit{robust accuracy} of a defense model is defined by their performance against the strongest adversarial attack within a specified $\ell_p$ boundary \cite{obfuscated_icml_2018}.
For RBF-CNN, we reconstruct images by drawing samples from the RBF filters. 
Noise is injected to the images during this sampling process using an isotropic Gaussian distribution (see Eqn. \ref{eq:sample}).
The RBF filters and the reconstruction step are otherwise differentiable.

We apply the following  wide range of adversarial attacks to evaluate the robustness of our models:

\begin{itemize}
	\item 
Static Attacks.
Adversarial examples are generated by removing the non-differentiable noise injection step from the reconstruction layer, i.e $\bm{z_j = \mu_j}$ in Eqn. \ref{eq:sample}.
Here, we consider single-step attack FGSM \cite{fgsm_iclr_2015}, iterative attacks such as PGD \cite{madry_iclr_2018}, MI-FGSM \cite{mifgsm_cvpr_2018}, CW \cite{cwAttack_sp_2017}, DAA \cite{daa_AAAI_2018}, EAD \cite{ead_aaai_2018}.
\smallskip
\item Adaptive Attacks. 
An adaptive attack is constructed after the defense model has been completely specified, such that the adversary can find the \textit{optimal} adversarial perturbations within the specified perturbation boundaries \cite{obfuscated_icml_2018}.

For our RBF-CNN models, we design the adaptive attack as a combination of BPDA and EoT (Expectation of Transformations) \cite{obfuscated_icml_2018}, as follows. 
In the forward propagation, we do not change anything in the network. 
In the backward propagation, we choose a differentiable approximation for the reconstruction layer, by considering $\bm{z_j = \mu_j}$ in Eqn. \ref{eq:sample} to efficiently compute the gradients.
Since the RBF filters are already differentiable, we do not need any approximation for the RBF layer.
We compute the expected loss using a Monte-Carlo method with $50$ simulations to find the strongest adversarial examples.
\smallskip
\item Black-box Attacks. We also evaluate against gradient-free SPSA attack \cite{spsa_icml_2018} to ensure that our framework is not giving any false sense of robustness using gradient-masking \cite{obfuscated_icml_2018}.
We apply \textit{black-box transfer attacks} where the adversarial examples are generated by attacking a standard CNN classifier with no defense.

\end{itemize}

\smallskip
\noindent \textbf{$\ell_{\infty}$ Bounded Attacks.}
Table \ref{table:mnist} presents the performance of our $rCNN$ and $rCNN_{+}$ models against $\ell_{\infty}$ bounded attacks. 
We choose the perturbation boundaries of $0.3$ and $0.031$ for MNIST and CIFAR-10 respectively.
We observe that for both MNIST and CIFAR-10, $rCNN_+$ achieves high robustness under $\ell_{\infty}$ perturbations bounds of $0.3$ and $0.031$ respectively.

\begin{table}[ht]
    \centering
    \resizebox{9.0cm}{!}{
        \begin{tabular}{l|cc|cc}
            \hline
            Attack Name    &  \multicolumn{2}{c|}{MNIST}   & \multicolumn{2}{c}{CIFAR-10} \\     \hline
            &    $rCNN$ & $rCNN_+$ & $rCNN$ & $rCNN_+$ \\ 
            &    \multicolumn{2}{c|}{($\ell_{\infty} \leq 0.3$)} & \multicolumn{2}{c}{($\ell_{\infty} \leq 0.031$)} \\ \hline
            Clean Test Data  & 99.6 & 99.5   & 85.0 & 85.1  \\ \hline
            {FGSM}  & 94.7 & 98.0   & 75.6 & 78.2  \\
            {PGD}   & 78.3 & 94.9   & 63.1 & 60.0  \\
            {MI-FGSM}   & 70.5 & 94.5&   65.8 & 70.2  \\
            {CW}   & 83.1 & 91.1  &  46.6 & 46.2  \\
            {DAA}   & 41.5 & 94.8  & 63.8 & 62.8  \\
            {BPDA+EoT}  & 39.4 & 88.4 &47.2 & 49.8  \\
            {SPSA} (Black-box)   & 61.4 & 92.9  & 62.4 & 67.6  \\
            FGSM (Black-box)  & 77.9 & 90.9   & 76.0 & 78.7  \\
            PGD  (Black-box) & 39.6 & 90.5& 80.3 & 81.8  \\
            CW (Black-box) & 63.0 & 95.5  & 76.5 & 79.4  \\ \hline
        \end{tabular}
    }
    \caption{Accuracy of  RBF-CNN models under $\ell_{\infty}$ attacks.}
    \label{table:mnist}
\end{table}

\begin{table}[t]
    \centering
    \resizebox{9.0cm}{!}{%
        \begin{tabular}{clcccc}
            \hline 
             & Defenses & \#Epochs & \multicolumn{1}{c}{Time/ Epoch} & Overhead & \multicolumn{1}{c}{Total} \\ \cline{2-6}
            \multirow{3}{*}{\begin{turn}{0}{\scriptsize MNIST} \end{turn}} & \begin{tabular}[c]{@{}l@{}}Madry  \end{tabular} & 85 & 112  & 0 &9,520   \\  
            & $rCNN$  & 100 & 4  &  600  & 1,000  \\
            & $rCNN_+$ &100 &11  & 900  & \textbf{2,000} \\ 
            \hline
            \multirow{3}{*}{\begin{turn}{0}{\scriptsize CIFAR-10}\end{turn}} & \begin{tabular}[c]{@{}l@{}}Madry  \end{tabular} & 205 & 1200  & 0 & 246,000  \\  
            & $rCNN$ & 600 & 29  & 2700  & 20,100 \\ 
            & $rCNN_+$  & 600 & 51   & 3600  & \textbf{34,200}  \\ \hline
        \end{tabular}
    }
    \caption{Training time comparison (in seconds).
    }
    \label{table:time_compare}
\end{table}

One main advantage of RBF-CNN compared to the adversarial training frameworks is that it significantly reduces the training time required. 
Here, we compare the training times of our RBF-CNN models with Madry's models, that achieved robustness only for $\ell_{\infty}$ perturbations \cite{madry_iclr_2018}.

Table \ref{table:time_compare} shows the training times when executed on a GTX 1080Ti GPU.
We observe that the training time of $rCNN_+$ models are more than $4.5\times$ and $7\times$ faster than the Madry models for MNIST and CIFAR-10 respectively.
Even as our proposed approach significantly reduces the training time required, our RBF-CNN models are still able to achieve similar performances as Madry's models for $\ell_{\infty}$ perturbations in both MNIST and CIFAR-10 as shown in Table \ref{table:exp_compare}. 
Also, our best RBF-CNN model for MNIST significantly outperforms the only successful manifold defense model by Schott et al. (2019) \cite{mnist_multiple_iclr2019}.
We  also compare with other input transformation \cite{thermometer_iclr_2018}\cite{transform_iclr_2018} and network randomization \cite{sap_iclr_2018} techniques to show that unlike these defenses, our RBF-CNN models remain robust against all type of attacks.

 \begin{table}
     \centering
     \resizebox{9.0cm}{!}{%
         \begin{tabular}{llcc}
             \hline 
             & Defenses & {\begin{tabular}[c]{@{}l@{}}Robust Acc.\end{tabular}} & {\begin{tabular}[c]{@{}l@{}}Strongest Attack\end{tabular}} \\ \cline{2-4}             
             & Baseline (no defense) & 0 & PGD \\
             \multirow{3}{*}{\begin{turn}{90}{\scriptsize MNIST} \end{turn}}
             & Madry et al., (2018) \cite{madry_iclr_2018} & \textbf{88.6} & DAA \\
             & Schott et al.,(2019) \cite{mnist_multiple_iclr2019} & 78.0 & Deep-Fool \\
             & $rCNN$  & 39.4 & BPDA+EoT \\
             & $\bm{rCNN}_+$ & \textbf{88.4} & BPDA+EoT \\            
             \hline 
             & Baseline (no defense) & 0 & PGD \\  
			 \multirow{7}{*}{\begin{turn}{90}{\scriptsize CIFAR-10} \end{turn}}              
             & Madry et al., (2018) \cite{madry_iclr_2018} & 44.7 & DAA \\               
             & Buckman et al., (2018) \cite{thermometer_iclr_2018} & 30 & BPDA \\
             & Ma et al., (2018) \cite{lid_iclr_2018} & 5 & CW \\
             & Dhillon et al., (2018) \cite{sap_iclr_2018} & 0 & EoT \\
             & Song et al., (2018) \cite{pixeldefend_iclr_2018} & 9 & BPDA \\
             & Dezfooli et al., (2019) \cite{cure_cvpr_2019} & 41.4 & PGD \\
             & $\bm{rCNN}$ & \textbf{46.6} & CW \\
             & $rCNN_+$  &  46.2 & CW \\
             \hline
         \end{tabular}
     }
     \caption{Comparison of robust accuracy against $\ell_{\infty}$ bounded adversarial attacks. 
     Perturbation boundaries for MNIST and CIFAR-10 are set to $0.3$ and $0.031$ respectively.}
     \label{table:exp_compare}    
\end{table}
Another limitation of adversarial training models is that they provide robustness only within a pre-specified boundary from where the adversarial examples were produced for their training, and offer no robustness guarantee slightly beyond these boundaries.
Table \ref{table: beyond} shows that  \cite{madry_iclr_2018} achieve high robust accuracies within their specified $\ell_{\infty}$ bounds of $0.3$ and $0.031$ for MNIST and CIFAR-10 respectively.
However, their robust accuracies drastically drop beyond those bounds when tested against PGD attack with $\ell_{\infty} \leq 0.35$ for MNIST  and at $\ell_{\infty} \leq 0.05$ for CIFAR-10.
In contrast, RBF-CNN models achieve significantly high accuracies at these $\ell_{\infty}$ boundaries.
\begin{table}
    \centering
    \resizebox{7.5cm}{!}{%
        \begin{tabular}{lcccc}
            \hline
            & \multicolumn{1}{l}{$\ell_{\infty}$-bounds} & Madry's & $rCNN$ & $rCNN_+$\\ \cline{2-5} 
            \multirow{2}{*}{MNIST} & 0.3 & \textbf{88.6} & 39.4 & \textbf{88.4}\\
            & 0.35 & 42.9$^{\dagger}$ & 10.5 & \textbf{75.8} \\ \hline
            \multirow{2}{*}{CIFAR-10} & 0.031 & 44.7 & \textbf{46.6} & 46.2 \\
            & 0.05 & 25.7$^{\dagger}$ & \textbf{37.1} & 34.7 \\ \hline
        \end{tabular}
    }
    \caption{Comparison of the robust accuracies at different 
        perturbation boundaries. $^{\dagger}$Evaluated only against PGD attacks.}
        \vspace*{0.1in}
    \label{table: beyond} 
\end{table}

\begin{figure}[htbp]
	\centering
	\includegraphics[width=0.6\linewidth,height=90pt]{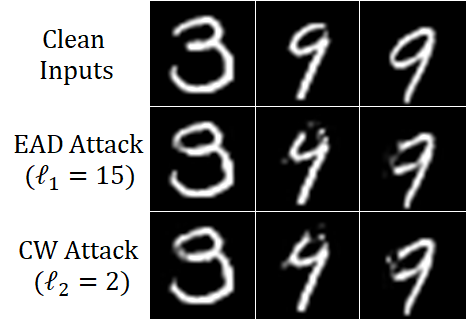}
	\captionof{figure}{Visual perception of MNIST images changed under $\ell_1$ and $\ell_2$ bounded attacks for RBF-CNN models.}
	\label{fig:l1l2}
\end{figure}

~\\ 
\textbf{$\ell_{1}$ and $\ell_2$ Bounded Attacks. }
For our experiments on $\ell_{1}$ and $\ell_2$ bounded attacks, we choose $\ell_1$=15 and $\ell_2$=2 for MNIST and $\ell_1$=20 and $\ell_2$=1 for CIFAR-10. 
For MNIST, we see that the perception of the adversarial images is changing around these bounds (see Fig. \ref{fig:l1l2}).
Note that, such behavior cannot be observed when attacking a non-robust classifier \cite{robustnessDrops_iclr_2018}.

\begin{table}[t]
    \centering
    \resizebox{9.0cm}{!}{
        \begin{tabular}{llccccc}
            \hline
            \multirow{2}{*}{$\ell_1$ Attacks} & \multirow{2}{*}{} & \multicolumn{2}{c}{Static} & \multicolumn{1}{c}{Adaptive} & \multicolumn{2}{c}{Black-box}\\
            &  & PGD & EAD & BPDA+EoT & PGD & EAD \\ \hline
            \multirow{3}{*}{\begin{tabular}[c]{@{}l@{}}MNIST\\ ($\ell_1 = 15$)\end{tabular}} & Madry & \textbf{77.5} & 90.4 & - & 97.9 & 98.2\\
            & $rCNN$  & 89.8 & \textbf{69.9}& 83.1 & 98.4 & 99.3 \\
            & $rCNN_{+}$   & 95.7 & \textbf{83.4} & 90.5 & 98.9 & 99.3 \\ \cline{2-7}
            \multirow{3}{*}{\begin{tabular}[c]{@{}l@{}}CIFAR-10\\ ($\ell_1 = 20$)\end{tabular}} & Madry & \textbf{34.5} & 35.5 & - & 86.2 & 86.1 \\
            & $rCNN$ & 69.2 & 61.0 & \textbf{55.5} & 82.0 & 81.1 \\
            & $rCNN_{+}$ & 76.1 & \textbf{61.9} & 66.8 & 82.9 & 82.3 \\ \hline
            
            \multirow{2}{*}{$\ell_2$ Attacks} & \multirow{2}{*}{Defenses} & \multicolumn{2}{c}{Static} & \multicolumn{1}{c}{Adaptive} & \multicolumn{2}{c}{Black-box}\\  
            &  & PGD & CW & BPDA+EoT & PGD & CW \\ \hline
            \multirow{3}{*}{\begin{tabular}[c]{@{}l@{}}MNIST\\ ($\ell_2 = 2$)\end{tabular}} & Madry & \textbf{81.9} & 91.0 & - & 95.6 & 97.2\\
            & $rCNN$ & 80.0 & 72.3 & \textbf{64.2} & 88.3 & 97.0 \\
            & $rCNN_{+}$ & 92.6 & \textbf{83.9} & 87.5 & 96.5 & 98.6 \\ \cline{2-7}
            \multirow{3}{*}{\begin{tabular}[c]{@{}l@{}}CIFAR-10\\ ($\ell_2 = 1$)\end{tabular}} & Madry & \textbf{28.1} & 43.3 & - & 85.2 & 85.1 \\
            & $rCNN$ & 62.1 & 47.9 & \textbf{45.2} & 79.1 & 76.8 \\
            & $rCNN_{+}$ & 67.3 & \textbf{54.5} & 54.8 & 81.1 & 79.5 \\ \hline
        \end{tabular}
    }
    \caption{Performance against $\ell_1$ and $\ell_2$ bounded attacks.}
    \label{table:l1l2}
\end{table}

Table \ref{table:l1l2} presents the performance of  RBF-CNNs against different $\ell_1$ and $\ell_2$ bounded attacks.
We observe that RBF-CNN models significantly outperform Madry's models, that are trained to provide robustness for $\ell_{\infty}$ perturbations.

~\\
\textbf{Certification for $\ell_2$-perturbations. }
Certification provides a lower bound guarantee of robust accuracy.
Since the reconstruction process of our RBF-CNN models mitigate \textit{any} minor perturbations, it also improves the classification performance against random Gaussian perturbed images.
We apply the existing certification technique of randomized smoothing defenses \cite{certifyNoise_icml_2019} to further demonstrate that our RBF-CNN models also improve the certified robustness for $\ell_2$ perturbations compared to the baseline models with no defense.

For each input ${\bm x}$, we sample $50,000$ noisy samples from an isotropic Gaussian distribution ${\bm \epsilon} \sim \mathcal{N}(0, \tau^2I)$.
For MNIST, we set $\tau$ to $0.2$ for the baseline and $rCNN$ and $0.3$ for $rCNN_+$.
For CIFAR-10, we set $\tau=0.1$ for all the models.
Table \ref{table:certify} shows the certified robust accuracies.
We see that for CIFAR-10, the robustness of the CNN classifier is improved by adding  RBF and reconstruction layers. 
The results for $rCNN_{+}$ suggests that the certified robustness of our models are further improved by augmenting minor noises the images for training.
To summarize, our RBF-CNN models improves the empirical robustness against any minor $\ell_{p\geq 1}$ perturbations as well as certified robustness for $\ell_2$ perturbations.
\begin{table}[t]
    \centering
    \resizebox{8cm}{!}{%
        \begin{tabular}{cc|ccc}
            \hline
            & & $\ell_2 = 0.5$ & $\ell_2 = 0.7$ & $\ell_2 = 1.0$ \\ \cline{3-5}
            \multirow{3}{*}{MNIST} & Baseline & 92.7 & 84.4 & 0.0 \\
            & $rCNN$ & 94.7 & 85.3 & 0.0 \\
            & $rCNN_+$ & \textbf{98.0} & \textbf{96.3} & \textbf{89.2} \\\hline
            &  & $\ell_2 = 0.25$ & $\ell_2 = 0.3$ & $\ell_2 = 0.35$ \\ \cline{3-5}
            \multirow{3}{*}{CIFAR-10} & Baseline  & 10.2 & 8.1 & 4.7 \\
            & $rCNN$ & 37.2 & 32.3 & 25.3 \\
            & $rCNN_+$ & \textbf{47.2} & \textbf{40.6} & \textbf{33.4} \\\hline
        \end{tabular}
    }
    \caption{Certified robust accuracy for $\ell_2$ perturbations.}
    \label{table:certify}
\end{table}
\subsection{Interpretable Loss Gradients}
\label{sec:visual_evidence}
\begin{figure}[htbp]
	\centering
	\includegraphics[width=1\linewidth,height=115pt]{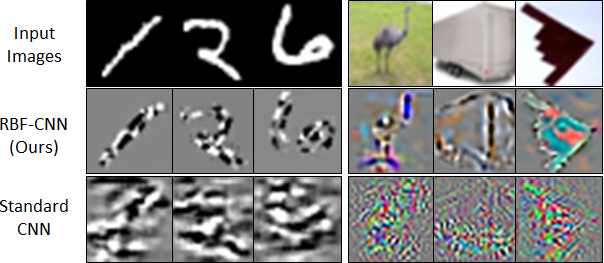}
	\caption{Visualization of loss gradients.}
	\label{fig:loss_grad}
\end{figure}

Figure \ref{fig:loss_grad} visualizes the loss gradients of our RBF-CNN, $rCNN_+$ versus standard CNN.
These loss gradients represent the most important pixels for the classifier.
All these loss gradients are obtained in one single step by computing the losses with respect to the input pixels.
For standard CNNs, these gradients appear noisy and incoherent patterns.
In contrast, the loss gradients for RBF-CNN are aligned with human perception without any pre-processing other than scaling and clipping.
This ensures that our RBF and reconstruction layers do not introduce gradient masking in our RBF-CNN framework.
Further, Etmann et al. (2019) \cite{saliency_icml_2019} demonstrate that only the robust classification models exhibit such interpretable saliency maps.

\begin{figure}[htbp]
	\centering
	\includegraphics[width=1\linewidth,height=140pt]{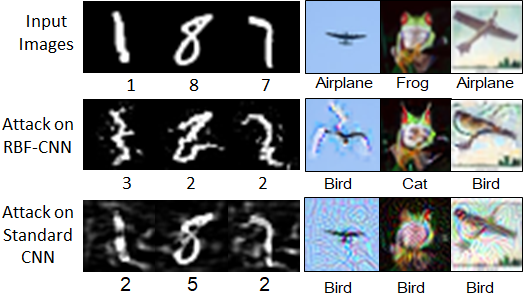}
	\caption{Adversarial Images generated using PGD attack.}
	\label{fig:visual_evidence}
\end{figure}

A first-order adversary iteratively uses these loss gradients to minimize the salient features of the original class and maximize the salient features of a different class to fool a model.
Thus, it is often not possible to fool a robust classifier as such perturbation may not exist within a small $\ell_p$ neighborhood and the adversary would require larger perturbations.
Consequently, as we allow a large $\ell_2$ boundary and apply the static PGD attacks, we observe sharp salient features of a different class appear in the generated images for our robust RBF-CNN models (see Fig. \ref{fig:visual_evidence}).
In contrast, the attack only able to produce a noisy version of the clean image for the non-robust standard CNN models.

\subsection{Robustness vs. Accuracy Trade-off at run-time}
Ideally, a user should be able to select an optimal trade-off between robustness or accuracy performance at run-time.
In RBF-CNN, we can vary the level of injected noise, $\beta_2$, during test time to achieve a different degree of robustness versus accuracy performance.
Table \ref{tab:noise} presents the performance of $rCNN_+$ as we vary  $\beta_2$.
We see that $rCNN_+$ is able to achieve robustness for all $\ell_1$, $\ell_2$ and $\ell_\infty$ bounded attacks even when we remove the randomization step by setting $\beta_2 = 0$.
Further, the robust accuracy doubles for CIFAR-10 when the noise level of  $\beta_2=1.75$ is injected with minimum effect on its clean data accuracy.
To the best of our knowledge, none of the existing methods allow such a trade-off between robustness versus accuracy at run-time.

\begin{table}[htbp]
    \centering
    \resizebox{8.5cm}{!}{%
        \begin{tabular}{l|ccc|ccc}
            \hline
            Attack Name  & \multicolumn{3}{c|}{MNIST}  & \multicolumn{3}{c}{CIFAR-10} \\  \hline
            $\beta_2$ & {$0.0$} & {$1.0$} & {$1.75$} & {$0.0$} & $1.5$ & $1.75$ \\ \hline
            Clean Test Data & 99.5 & 99.5 & 99.4 &   89.2 & 87.2 & 85.1 \\ \hline
            & \multicolumn{3}{c}{$\ell_{\infty} \leq 0.3$}  & \multicolumn{3}{c}{$\ell_{\infty} \leq 0.031$} \\  \cline{2-7}
            PGD & 91.6 & 93.1 & 94.9 & 51.1 & 59.4 & 60.0 \\
            CW & \textbf{77.7} & \textbf{85.8} & 91.1   & \textbf{20.9} & \textbf{44.3} & \textbf{46.2} \\
            BPDA+EoT & - & 88.4 & \textbf{88.4} & - & 50.6 & 49.8 \\
            PGD (black-box) & 93.7 & 92.3 & 90.5 & 86.6 & 85.1 & 81.8 \\ \hline
            & \multicolumn{3}{c}{$\ell_{1} \leq 15$} & \multicolumn{3}{c}{$\ell_{1} \leq 20$} \\ \cline{2-7}
            PGD & 93.8 & 94.6 & 95.7 & 73.3 & 76.2 & 76.1 \\
            EAD & \textbf{80.6} & \textbf{82.0} & \textbf{83.4} &  \textbf{41.1} & \textbf{61.1} & \textbf{61.9} \\
            BPDA+EoT & - & 90.2 & 90.5 &  - & 67.9 & 66.8 \\ \hline
            & \multicolumn{3}{c}{$\ell_{2} \leq 2$} & \multicolumn{3}{c}{$\ell_{2} \leq 1$} \\ \cline{2-7}
            PGD & 88.4 & 90.5 & 92.6   & 63.8 & 66.8 & 67.3 \\
            CW & \textbf{80.4} & \textbf{82.2} & \textbf{83.9}  & 36.5 & \textbf{52.7} & \textbf{54.5} \\
            BPDA+EoT & - & 87.1 & 87.5  & - & 56.4 & 54.8 \\ \hline
        \end{tabular}%
    }
    \caption{Robustness versus Accuracy tread-off at run-time for RBF-CNN models, $rCNN_+$ by varying the noise level hyper-parameter, $\beta_2$ in the reconstruction layer.}
    \label{tab:noise}
\end{table}

\section{Conclusion}
Existing successful defense models typically achieve robustness only for a specific perturbation type while providing no guarantee for other perturbation types.
Towards this, we presented an ``approximate manifold'' defense called RBF-CNN that achieves robustness for all minor perturbations in any $\ell_p$-norm with $p\geq 1$.
Our RBF-CNN utilizes an RBF and a reconstruction layer.
We propose to capture the density of small image patches, instead of capturing the complete data manifold to address the limitations of the existing manifold defenses.
Our experimental results on MNIST and CIFAR-10 demonstrate that we can train a single RBF-CNN model to provide robustness for all $\ell_1$, $\ell_2$, and $\ell_{\infty}$ perturbations.
While previously, the success of the only effective manifold-based defense remains limited to MNIST \cite{mnist_multiple_iclr2019}, we achieve robustness for a much complex image dataset, called CIFAR-10.

Even though our proposed RBF-CNN models achieve robustness against minor additive perturbations in-terms of any $\ell_{p\geq 1}$ norms, we do not necessarily provide any guarantee for other perturbation types such as spatial \cite{spatialAttack_icml_2018,blindspotAttack_iclr_2018,wassersteinAttack_icml_2019} or color transformations \cite{functionalAttack_nips_2019} or naturally occurring common perturbation types \cite{commonPerturbation_iclr_2019,commonPerturbation_2_iclr_2019}.
While it is crucial to develop universally robust defense models for real-world applications to improve their reliability, it remains an open problem to the AI community.

\section*{Acknowledgment}
This research is supported by the National Research Foundation Singapore under its AI Singapore Programme (Award Number: AISG-RP-2018-008).

\bibliographystyle{IEEEtran}
\bibliography{IEEEabrv,robustness_bib}
\end{document}